\documentclass[superscriptaddress,preprint]{revtex4}
\usepackage{mathrsfs}
\usepackage{amssymb}
\usepackage[tbtags]{amsmath}
\usepackage{graphicx}
\usepackage{epsfig,graphicx,times}
\usepackage{color}
\usepackage{subfigure}
\usepackage{overpic}

\begin{document}

\title{Microwave transmissions through superconducting
coplanar waveguide resonators with different coupling
configurations}
\author{ZHANG Si-Lei}
\affiliation{Quantum Optoelectronics Laboratory, 
Southwest Jiaotong University, Chengdu  610031, China}
\affiliation{Institute of Electromagnetic Field and Microwave
Technology, Southwest Jiaotong University, Chengdu 610031, China}
\author{LI Hai-Jie}
\affiliation{Quantum Optoelectronics Laboratory, 
Southwest Jiaotong University, Chengdu  610031, China}
\author{L. F. Wei\footnote{weilianfu@gmail.com}}
\affiliation{Quantum Optoelectronics Laboratory, 
Southwest Jiaotong University, Chengdu  610031, China}
\affiliation{State Key Laboratory of Optoelectronic Materials Technologies, 
Sun Yat-sen University, Guangzhou 510275, China}

\author{FANG Yu-Rong}
\affiliation{Research Institute of Superconductor Electronics,
Nanjing University, Nanjing 210093, China}
\author{WANG Yi-Wen}
\affiliation{Quantum Optoelectronics Laboratory, Southwest Jiaotong
University, Chengdu 610031, China}
\author{ZHOU Pin-Jia}
\affiliation{Quantum Optoelectronics Laboratory, Southwest Jiaotong
University, Chengdu 610031, China}
\author{WEI Qiang}
\affiliation{Quantum Optoelectronics Laboratory, Southwest Jiaotong
University, Chengdu 610031, China}
\author{CAO Chun-Hai}
\affiliation{Research Institute of Superconductor Electronics,
Nanjing University, Nanjing 210093, China}
\author{XIONG Xiang-Zheng}
\affiliation{Institute of Electromagnetic Field and Microwave Technology,
Southwest Jiaotong University, Chengdu 610031, China}

\date{\today}

\begin{abstract}

We design and fabricate two types of
superconducting niobium coplanar waveguide microwave resonators with
different coupling capacitors on high purity Si substrates. Their
microwave transmissions are measured at the temperatures of $20$ mK.
It is found that these two types of resonators possess
significantly-different loaded quality factors; one is
$5.6\times{10}^{3}$, and the other is $4.0\times{10}^{4}$. The
measured data are fitted well by classical ABCD matrix approach. We
found that the transmission peak deviates from the standard
Lorentizian with a frequency broadening.

PACS number(s): 84.40.Az, 74.78.Na, 85.25.Am
\end{abstract}

\maketitle

In recent years much research attention has been paid to
superconducting coplanar waveguide (SCPW) resonators due to their
easy fabrications and many potential applications. A typical example
is for the kinetic inductance detectors\cite{R1}, which can work in
optical-, UV-, and X-ray ranges\cite{R1,R2,R3,R4,R5}. Other
applications include bifurcation- \cite{R6} and parametric
amplification\cite{R7,R8,R9}, as well as solid-state quantum
computation\cite{R10,zm}.
Technologically, CPW resonators can be easily fabricated due to
their simple single-layer film structures: Their center strips and
the relevant grounds are made in the same planes with the same
material. Therefore, they are very convenient to be integrated into
various parallel devices.

Usually, quasi-TEM modes propagate along the CPWs with
sufficiently-narrow center strips. For the practical applications,
two types of CPW resonators are mainly considered. The $\lambda/4$
shorted CPW resonators, one side is coupled to the feed line for
measurement and the other side is shorted directly, are mainly used
for kinetic inductance detectors. While, the $\lambda/2$ opened CPW
resonators are mainly used in quantum information, solid-state
quantum optics, and parametric amplifiers, etc.. In these
structures, the resonators can be coupled outsides via various
capacitive configurations, such as finger, gap\cite{R11}, and
overlap ones, respectively.
In this paper, we experimentally demonstrated the $\lambda/2$ opened
Nb SCPW resonators with two types of coupled capacitors. By using
the standard vector network analyzer we measured their microwave
transmission coefficients $S_{21}$ at low temperature 20 mK and
obtained very different transport properties, which should be useful
for the further applications.

\begin{center}
\begin{overpic}[width=8.5cm]{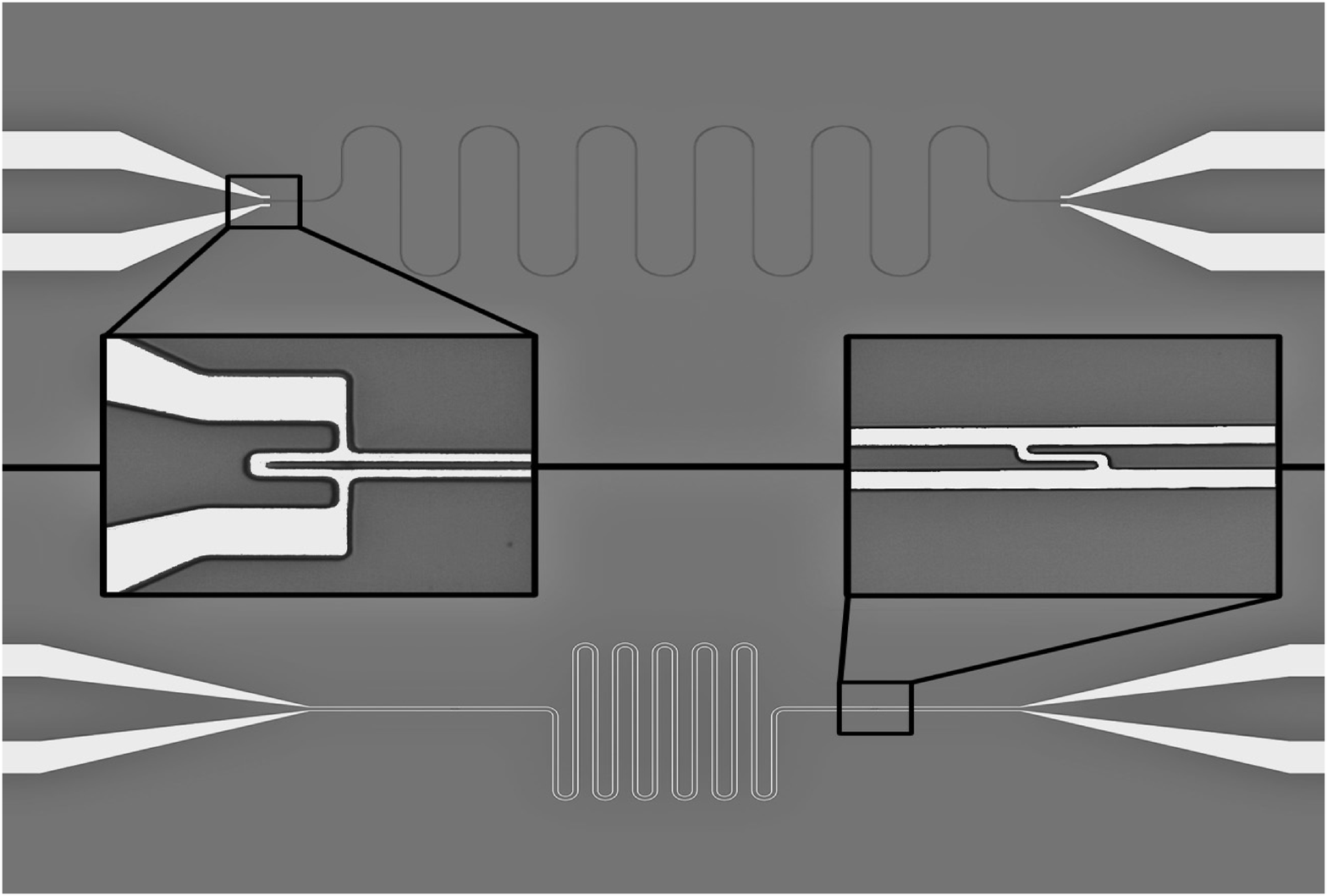}
\put(10,57){\bf a} \put(10,38){\bf b} \put(10,9){\bf c}
\put(80,38){\bf d}
\end{overpic}
\end{center}
{\small\it Fig.~1. The demonstrations of $\lambda/2$ Nb CPW
resonators with different coupling configurations. (a) The resonator
couples outside via two overlap capacitors shown in (b). (c) The
resonator couples outside via two finger capacitors shown in (d).
Here, the white region represents the applied Si substrates, and the
grey region is the deposited metallization on the substrate}.

The two types of Nb $\lambda/2$ CPW resonators designed and
fabricated are shown in Fig.~1. Where, both of the resonators were
fabricated on a high-purity Si wafer. The thickness of the Nb films
deposited (by sputtering method) on the Si substrate is about
$0.16\,\mu$m, and then the desired structures are produced by the
photolithographic technology. For the first CPW resonator (a), the
center conductor has a width of 5$\mu$m, the gap between the center
conductor and ground plane is also 5$\mu$m, and the length of the
meandering resonator is $L_1=19.088$mm. The length of the coupling
finger capacitors (b) at input/output ports is $L_f=35\mu$m, and the
width $d_1$ of finger and the gap $s_1$ of the center conductor
separated from ground plans is the same as the width $w_1$ of the
center conductor, i.e., $s_1=d_1=w_1=5\mu$m. While, the second CPW
resonator shown in Fig.~1(c) has the relevant parameters:
$L_2=18.258$mm, $w_2=20\mu$m, $s_2=15\mu$m, $d_2=7.5\mu$m, the gap
between the two fingers is $5\mu$m, and the length of finger is
50$\mu$m.

The microwave transmissions through the resonators are measured in a
${}^3\!{He}$-${}^4\!{He}$ dilution refrigerator at the base
temperature of 20mK. We use a 14GHz Agilent E5071C vector network
analyzer (VNA) to measure the transmitted coefficient $S_{21}$ (see
Fig.~2). In order to avoid various non-linear effects, the two
resonators are driven respectively by -60$dBm$ and -50$dBm$ input
power. Theoretically, the fundamental frequency $f$ of a $\lambda/2$
resonator can be evaluated as
\begin{equation}
f=\frac{c}{\sqrt{\varepsilon_{eff}}}\frac{1}{2L},
\end{equation}
where $c$ is speed of electromagnetic wave in vacuum,
$\varepsilon_{eff}$ is the effective permittivity (which is the
function of the CPW geometry and the relative permittivity
$\varepsilon_r$ of substrate). And, $L=\lambda/2$ is the length of
the resonator.

\begin{center}
\begin{overpic}[width=8.5cm]{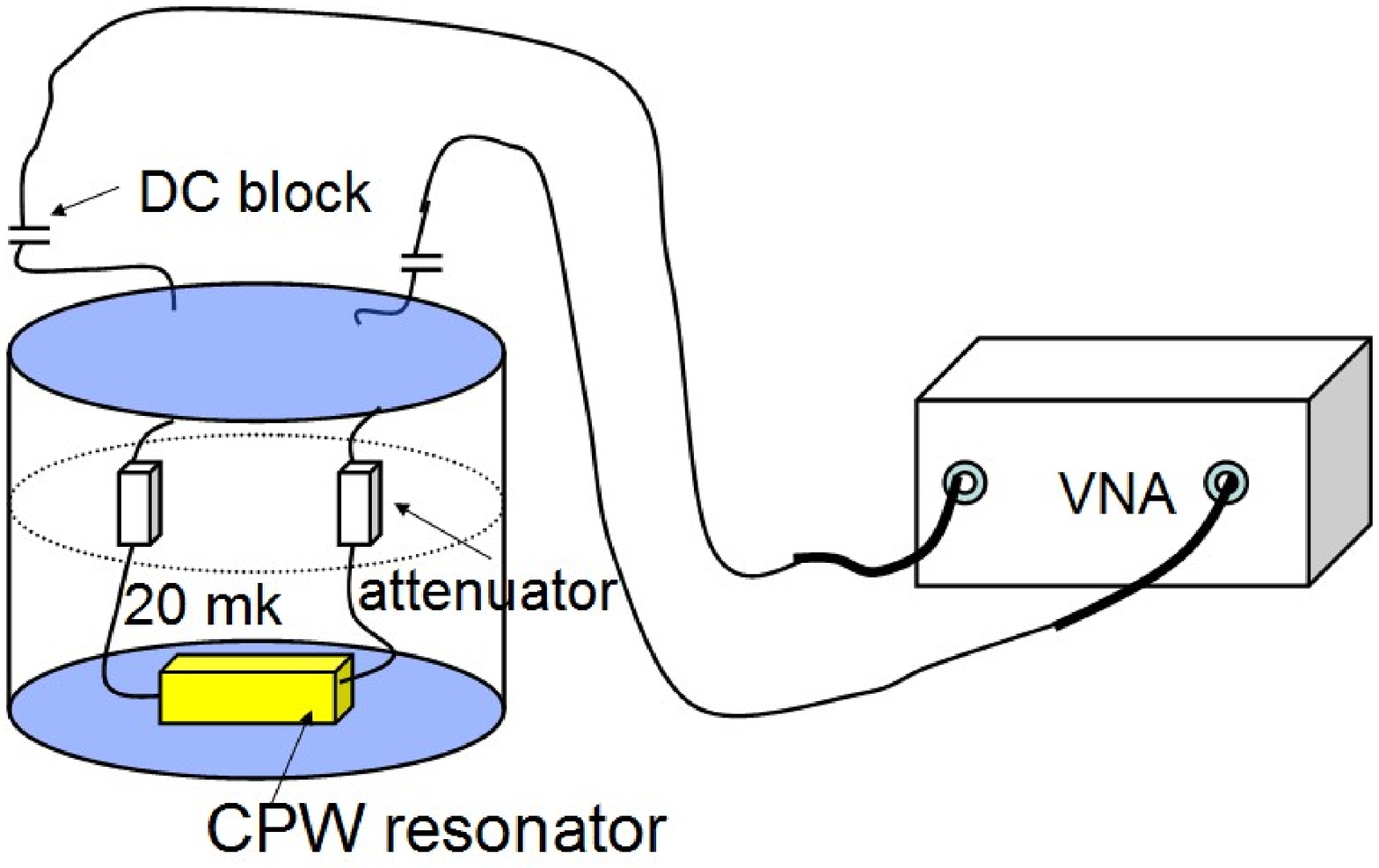}
\end{overpic}
\end{center}
{\small\it Fig.~2. Schematic diagram of the our low-temperature
microwave measurement system. Gray boxes represent the attenuators.}

The loaded quality factory $Q_L$ of the resonator can be obtained by
measuring its transmission coefficient $S_{21}$ (dB), i.e.,
$Q_L=f_0/\Delta f$ with $f_0$ is the measured resonant frequency and
$\Delta f$ the 3dB bandwidth of the transmitted peak. Physically,
the quality factory $Q_L$ can also be expressed by
\begin{equation}
\frac{1}{Q_L}=\frac{1}{Q_{ext}}+\frac{1}{Q_{int}},
\end{equation}
with
\begin{equation}
Q_{ext}=\frac{\omega_0 R'
C}{2},\,R'=\frac{1+\omega_0^2C_c^2Z_L^2}{\omega_0^2C_c^2Z_L},\omega_0=2\pi
f_0,
\end{equation}
describing the influence from the outsides\cite{R12}, and
\begin{equation}
Q_{int}=\frac{\pi}{2\alpha L},
\end{equation}
the internal losses. Above, $C_c,\,\alpha$ is the coupling
capacitance and the internal losses of the resonator, respectively.
Also, $Z_L=50\Omega$ is the characteristic impedance of outside the
microwave lines. Experimentally, the coupling coefficient
$g=Q_{int}/Q_{ext}$, between the resonator and the microwave feed
line\cite{R13}, can be determined by observing the insertion loss of
the resonator (i.e., the declination of peak from unity)\cite{R12}:
\begin{equation}
L_0=-20\log(\frac{g}{g+1}).
\end{equation}
\\
Here, $g=1,\,g>1$, and $g<1$ refer to critically coupled,
overcoupled, and undercoupled, respectively. Our measurements show
that both of the two CPW resonators measured here are undercoupled.
Furthermore, from Eqs.~(2) and (5) and also the measured $Q_L$ we
can deliver the $Q_{int}$- and $Q_{ext}$ parameters. Next, we can
calculate the coupling capacitance $C_c$ and the internal loss
$\alpha$.

\begin{center}
\begin{overpic}[width=8.5cm]{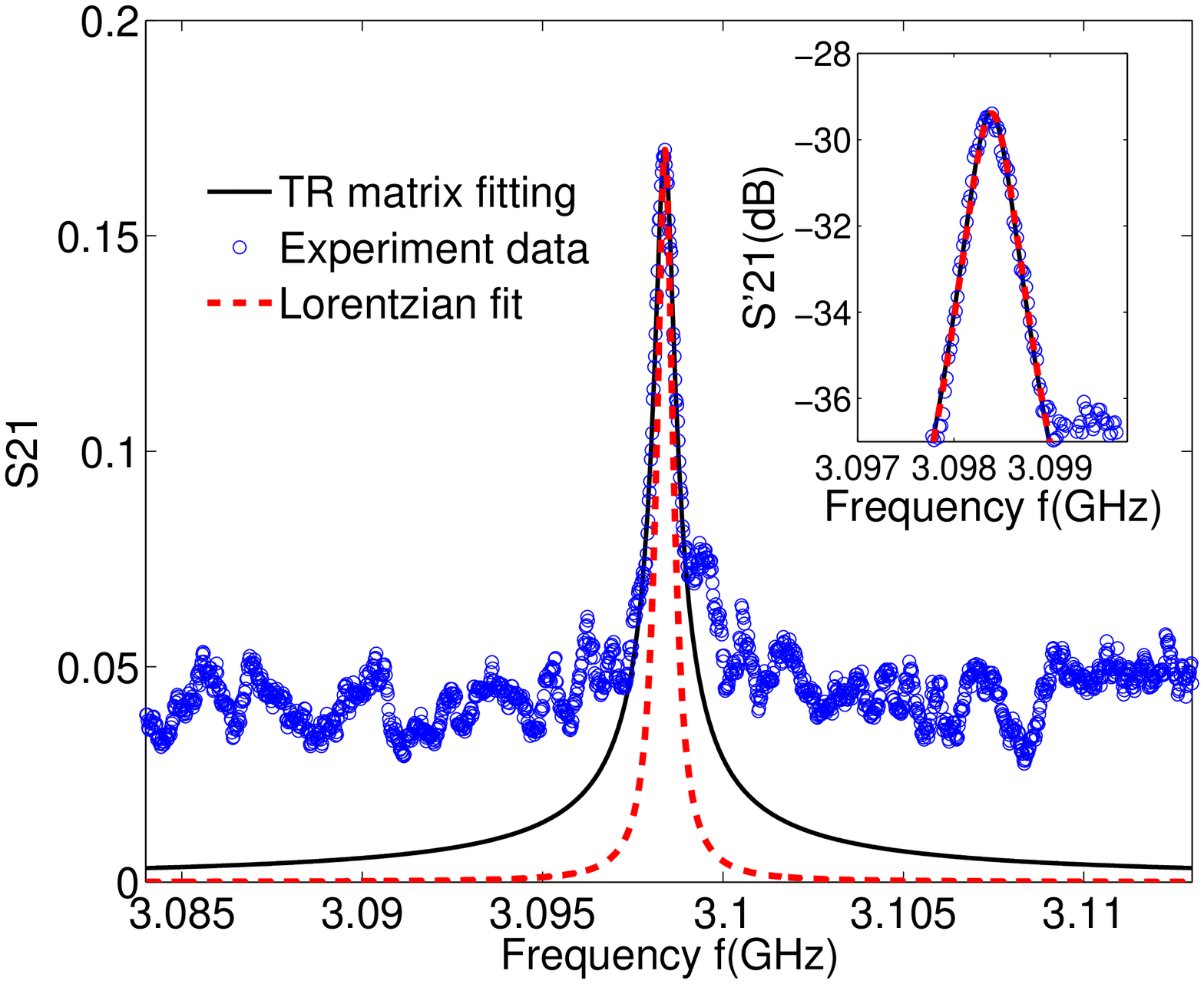}
\put(30,68){\bf (a)} \put(80,68){\bf (b)  }
\end{overpic}
\end{center}
\begin{center}
\begin{overpic}[width=8.5cm]{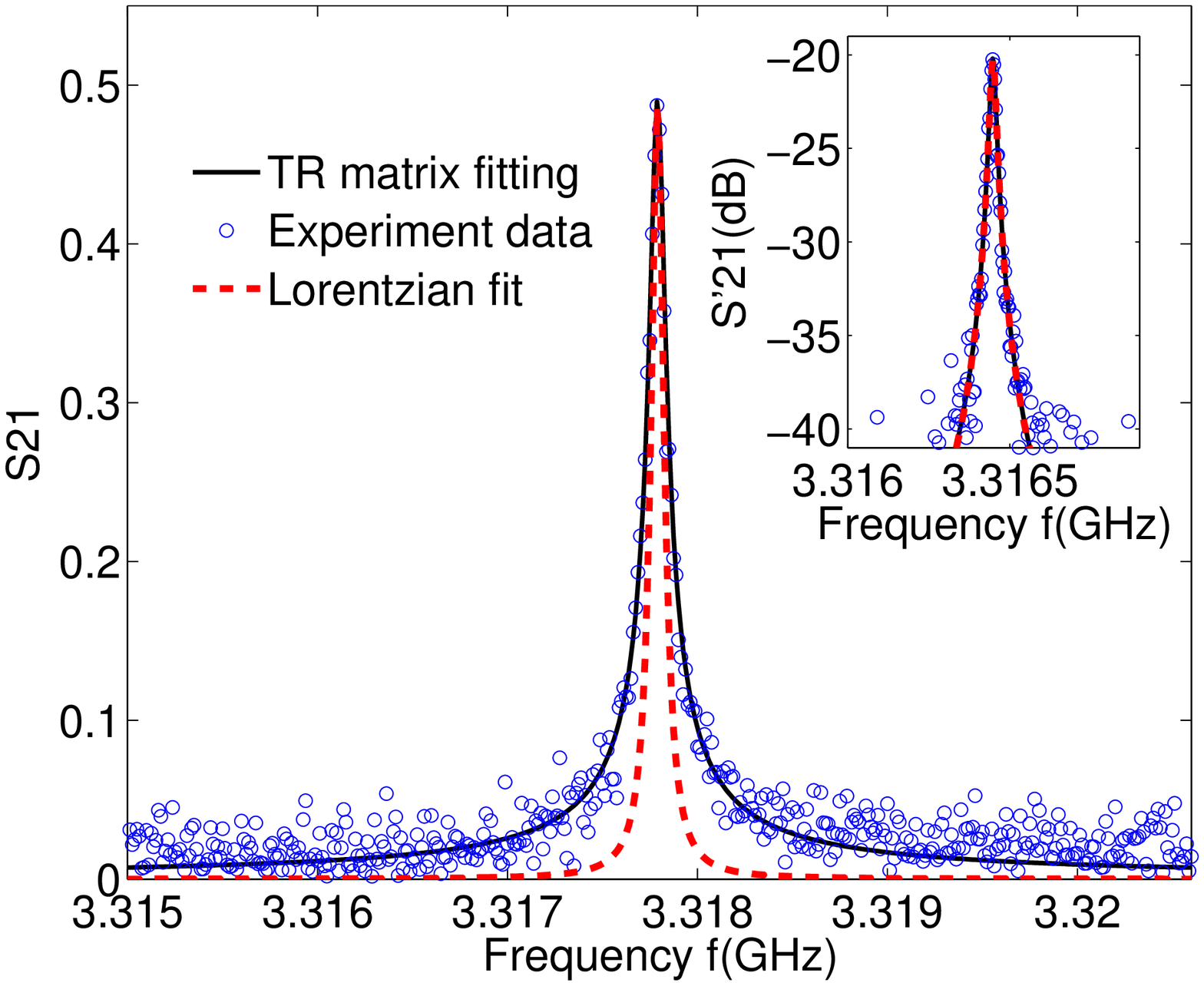}
\put(30,68){\bf (c)} \put(80,68){\bf (d)}
\end{overpic}
\end{center}
{\small\it Fig.~3. Amplitude of the measured transmission
coefficients $S_{21}$ through: (a) the first $\lambda/2$ resonator
and its dB format (b); (c) the second $\lambda/2$ resonator and its
dB format (d). Here, the blue points are the measured data, and the
black and red curves are fitted results by using transmission matrix
(TR) theory and Lorentzian line shape. The data indicate the
manifest background noise at the points deviated obviously from the
resonance}.

The measured $S_{21}$ parameter can be fitted by the usual
transmission ABCD matrix theory\cite{R13}, i.e.,
\begin{equation}
S_{21}=\frac{2}{A+B/Z_L+CZ_L+D}.
\end{equation}
Here, the ABCD parameters are determined by the equation
\begin{eqnarray}
  \left(\begin{array}{cc}
  A & B \\
  C & D
  \end{array}\right)=T_zT_{cpw}T_z,
  \end{eqnarray}
with
\begin{eqnarray}
  T_z=\left(\begin{array}{cc}

  1 & (j\omega C_c)^{-1} \\
  0 & 1
  \end{array}\right),\,\omega=2\pi f,
  \end{eqnarray}
and
  \begin{eqnarray}
  T_{cpw}=\left(\begin{array}{cc}
  \cos(\beta L) & jZ_{\rm cpw}\sin(\beta L) \\
  j(Z_{\rm cpw})^{-1}\sin(\beta L) & \cos(\beta L)
  \end{array}\right).
\end{eqnarray}
Where, $j$ is the imaginary unit, $\beta=\omega/v_p$,
$v_p=1/{\sqrt{L_lC_l}}$, and $Z_{\rm cpw}=\sqrt{L_l/C_l}$ is the
characteristic impedance with $L_l,\,C_l$ being the inductance and
capacitance per unit, respectively. Note that, for the present case
CPW has non-zero loss and thus $\beta$ in Eq.~(6) should be replaced
by $(\alpha+j\beta)/j$. Practically, $L_l$ and $C_l$ can be
calculated by using the so-called conformal
techniques\cite{R14,R15}:
\begin{equation}
L_l=\frac{\mu_0}{4}\frac{K(k')}{K(k)},\,C_l=4\varepsilon_0\varepsilon_{eff}\frac{K(k)}{K(k')},
\end{equation}
where $\mu_0=4\pi\times 10^{-7}$ H/m and $\varepsilon_0$
=8.85$\times$$ {10}^{-12}$ F/m are the permeability and permittivity
of free space, respectively. Also,
$\varepsilon_{eff}=(\varepsilon_r+1)/2$  with $\varepsilon_r=11.9$
for the Si substrate. And, $K(k)$ represents the first kind complete
elliptic integral with
\begin{equation}
k=\frac{w}{w+2s},\,k'=\sqrt{1-{k}^2}.
\end{equation}
With the above relations and the experimental data, we obtained
$C_l\approx1.46\times10^{-10}$ F/m for the first resonator, and
$C_l\approx1.58\times 10^{-10}$ F/m for the second resonator.

Take into account cable losses (which is measured as 14dB), the
$L_0$ in equation (5) is practically not the real insertion loss of
the CPW resonator. We can also change Eq.~(6) into the usual
logarithmic units(dB), i.e., $S_{21}'=20\log_{10}|S_{21}|$(dB).
Consequently, $C_c$ for each resonators can also be fitted by using
the least-square method, and the results are listed in Table I. It
is shown that, the fitted and the above calculation based Eq.~(3) of
the coupling capacitance $C_c$ agree well; the deviations $\delta
C_c=C'_c-C_c$ of the fitted $C'_c$ compared to the theoretical
calculations $C_c$ are really samll, e.g., $\delta C_c=-0.14\%$ for
the first resonator and $\delta C_c=-0.5\%$ for the second
resonator.

\begin{table}[h]
\caption{Main parameters of the two CPW resonators: $f_0$ and $Q_L$
are the measured resonant frequency and the loaded quality factor,
respectively. $C_c$ is the fitted coupling capacitance.}
\begin{tabular}{cccc}
\multicolumn{4}{c}{}\\
\hline Type\,\,\,\,\,\,\,\,\,\,\,\,\,\,\,\,\,\,&
$f_0$(GHz)\,\,\,\,\,\,\,\,\,\,\,\,\,\,\,\,\,\,
&$C_c$(fF)\,\,\,\,\,\,\,\,\,\,\,\,\,\,\,\,\,\,&$Q_L$
\\\hline
1\,\,\,\,\,\,\,\,\,\,\,\,\,\,\,\,\,\,&
3.09839\,\,\,\,\,\,\,\,\,\,\,\,\,\,\,\,\,\,
&$4.66$  \,\,\,\,\,\,\,\,\,\,\,\,\,\,\,\,\,\,&$5.6\times10^3$ \\
2 \,\,\,\,\,\,\,\,\,\,\,\,\,\,\,\,\,\,& 3.31779
\,\,\,\,\,\,\,\,\,\,\,\,\,\,\,\,\,\,
&$2.90$\,\,\,\,\,\,\,\,\,\,\,\,\,\,\,\,\,\,
&$4.0\times10^4$\\
\hline
\end{tabular}
\end{table}

Finally, we fit the peak shape of the measured $S_{21}$ parameters
by Lorentizian function
\begin{equation}
S_{21}(\omega)=A_0\frac{(\omega_0/2Q_L)^2
                           }{(\omega-\omega_0)^2+(\omega_0/2Q_L)^2},
\end{equation}
with $A_0$ being the measured value of the $S_{21}$ parameters at
their the resonance points. Physically, $A_0$ represents how much
power transport from one port to another in percentage so that we
must minus the cable losses and insertion loss. Certainly, $A_0=1$
refers to the full power transmission. It is seen from Fig.~3 that,
the measured peak shape deviates the standard Lorentizian shape with
certain frequency broadenings.

In conclusion, we have designed and fabricated two types of Nb
$\lambda/2$ symmetrically coupled CPW resonators and measured their
microwave transmission properties at the temperature of 20 mk for
lower input powers. The measured data around the resonant points
have been fitted well by using the usual ABCD matrix model and the
relevant parameter such as the coupling capacitance $C_c$ of the CPW
resonators were determined. It was found that the measured
transmission peak were not the standard Lorentizian shape. The
reason of the found frequency broadenings would be investigated in
detail in future. Finally, we found that the background noise is
dominant at the non-resonant points, which should be further
suppressed by using the further filtering technique.

{\bf Acknowledgements}: This work was Supported by the National Natural Science Foundation of China under Grant Nos. 11174373, 11204249, and the National
Fundamental Research Program of China, through Grant No.
2010CB923104. We thank Profs. Yu Yang
and Wu Pei-Heng for kindly help and discussions.


\end{document}